\input ./style/arxiv-general.cfg
\documentclass[MSNbibl,nameyear,dvips]{arxstspdf}
\makeatletter
   \@ifpackageloaded{graphicx}{}{\usepackage{graphicx}}
\makeatother
\usepackage{flushend}
\usepackage{stfloats}
\volume{30}
\issue{4}
\pubyear{2015}
\firstpage{582}
\lastpage{596}
\doi{10.1214/15-STS528}% Updated by VTEXPTS2LaTeX.exe, 04.09.2015 15:31
\docsubty{FLA}

\makeatletter
\setattribute{abstract}    {skip} {20\p@}
\setattribute{keyword}     {skip} {8\p@}
\setattribute{frontmatter} {skip} {0\p@ plus 3\p@ minus 3\p@}
\setattribute{frontmatter} {cmd}  {}
\setattribute{abstract}   {width}  {35.5pc}
\setattribute{keyword}    {width}  {35.5pc}

\makeatother

\begin{document}
\begin{frontmatter}

\title{A Conversation with Nan Laird}%\thanksref{T1}
% kai straipsnis turi susijusiu diskusiju ir rejoinder'iu
%\relateddois{T1}{Discussed in \relateddoi{d}{10.1214/00-STSXXX} ...; rejoinder at \relateddoi{r}{10.1214/00-STSXXXX}.}
\runtitle{A Conversation with Nan Laird}
%\pdftitle{}

\begin{aug}
\author{\fnms{Louise}~\snm{Ryan}\corref{}\ead[label=e1]{louise.m.ryan@uts.edu.au}}
\runauthor{L. Ryan}
\affiliation{University of Technology Sydney}

\address{Louise Ryan is
Distinguished Professor of Statistics at University of Technology Sydney,
Australian Research Council Centre of Excellence in Mathematical and Statistical Frontiers,
UTS School of Mathematical and Physical Sciences, PO Box 123, Broadway, NSW 2007, Australia \printead{e1}.}
\end{aug}

% ABSTRACT
\begin{abstract}
Nan McKenzie Laird is the Harvey V. Fineberg Professor
of Biostatistics at the Harvard T.~H. Chan School of Public Health. She has
made fundamental contributions to statistical methods for longitudinal data
analysis, missing data and meta-analysis. In addition, she is widely known
for her work in statistical genetics and in statistical methods for
psychiatric epidemiology. Her 1977 paper with Dempster and Rubin on the EM
algorithm is among the top 100 most highly cited papers in science
[\textit{Nature} \textbf{524} (2014) 550--553]. Her applied work on medical practice errors is
widely cited among the medical malpractice community.

Nan was born in Gainesville, Florida, in 1943. Shortly thereafter, her
parents Angus McKenzie Laird and Myra Adelia Doyle, moved to Tallahassee,
Florida, with Nan and her sister Victoria Mell. Nan started college at Rice
University in 1961, but then transferred to the University of Georgia where
she received a B.S. in Statistics in 1969 and was elected to Phi Beta
Kappa. After graduation Nan worked at the Massachusetts Institute of
Technology Draper Laboratories where she worked on Kalman filtering for the
Apollo Man to the Moon Program. She enrolled in the Statistics Department
at Harvard University in 1971 and received her Ph.D. in 1975. She joined
the faculty of Harvard School of Public Health upon receiving her Ph.D., and
remains there as research professor, after her retirement in 2015.

In the 40 years that Nan M. Laird spent at Harvard she authored or
co-authored over 300 papers and three books, and mentored numerous graduate
students, postdoctoral fellows and junior faculty. According to Google
scholar, her work has been cited over 111,000 times. She has received many
awards for her varied contributions to statistical science. Some highlights
include the Samuel S. Wilks Award, her election as Fellow of the American
Association for the Advancement of Science, her election as Fellow of the
American Statistical Association, the Janet Norwood Prize, the F.~N. David
Award and, most recently, the Marvin Zelen Leadership Award in Statistical
Science. Professor Laird has served on many panels and editorial boards
including a National Academy of Science Panel on Airliner Cabin
Environment, which led to the elimination of smoking on airplanes.
Professor Laird chaired the Department of Biostatistics at the Harvard
School of Public Health from 1990--1999 where she was the Henry Pickering
Walcott Professor of Biostatistics. In this interview she talks about her
career, including her passion for mentoring students. She offers some
helpful advice about balancing work and family life, acknowledging the
powerful encouragement and support that her own family has given her over
the years.

The interview was conducted in Boston, Massachusetts, in July 2014. A~link
to Nan's full CV can be found at
\surl{www.hsph.harvard.edu/nan-laird/}.
\end{abstract}

% KEYWORDS
% Pirmas kwd is didziosios raides
\end{frontmatter}

%s1 #&#
\section{Early Life}\label{sec1}

\textit{Ryan}: Tell us about your early life.

\textit{Laird}: I~had a conventional southern childhood in Tallahassee,
Florida. My mother was a school teacher, who stayed home to care for my
sister and myself. My father was briefly a professor of political science,
but turned to state government and politics shortly after my sister and I~were born. Much of my father's job was stressful and not so interesting to
him; most of his energy went to buying and improving small parcels of rural
land. Our life centered around the church and family, gardening and my
father's projects. We lived in a small neighborhood in Tallahassee, full of
kids, and I~recall spending our days roaming free on the heavily wooded
local golf course. We had a great deal of independence. My childhood was a
happy time and I~am still close to many of my childhood neighbors and
friends.

\textit{Ryan}: Were you interested in math as a kid?

\textit{Laird}: I~always loved math and in high school it was my favorite
subject. My sister and I~went to the local public high school and some of
the teachers were outstanding. My choice of undergraduate college, Rice
University in Houston, was a mismatch. At the time it was a bit of a
backwater; women were not encouraged in mathematics nor in the sciences in
general. Women were such a minority that I~was always the only girl in my
math section. I~switched my major to French, but then I~got married to a
fellow Rice student, dropped out after my junior year, moved to New York
City and had my first child, Richard.

We moved to Georgia a few years later, and I~went back to undergraduate
studies at the University of Georgia. I~wanted to study something
practical, so I~decided on computer science. This was in the late sixties,
and computers were just becoming mainstream. Fortunately Computer Science
and Statistics were in the same department and the department chair, Carl
Kossack, told me that Statistics was so much more interesting, so I~studied
statistics. I~took a course from Kossack in Statistical Decision Theory
that used Herman Chernoff's book. I~loved this course! What appealed to me
was the idea that one could use math formulas to make mundane life
decisions, like whether or not to take your umbrella to work. Of course, in
retrospect I~realize the course was really about decision theory and not
much about statistics at all, but still fascinating. It was still one of
the experiences that persuaded me to major in statistics.

\textit{Ryan}: What did you do at first when you graduated from college and
what motivated you to go to graduate school? I~understand that at some
point you had a brief stint as a fashion model?

\textit{Laird}: Yes, I~did briefly try modeling and it was a lot of fun!
But I~figured it was not a good long-term career path. After graduating
from University of Georgia with a B.S. in statistics, we moved to Boston and
I~worked at MIT with the impressive title of ``engineer.'' It was actually
a pretty exciting project at MIT's Draper Labs working on technology to
support the Apollo Moon Program. My role was to develop computer programs
to test the Kalman filtering used for the inertial guidance system. The
main thing I~learned was that to work in a creative environment I~needed
more education, so after a couple of years, I~went to grad school in
statistics at Harvard.

%s2 #&#
\section{EARLY Days at Harvard}
\label{sec2}

\textit{Ryan}: What was it like being a graduate student in the Harvard
statistics department in the 1970s?

\textit{Laird}: I~really loved my years in graduate school. I~had great
colleagues among my fellow students, and I~always felt that I~was treated
well. I~did hear stories about the lack of women faculty in the department
and I~also heard a rumor that I~was not offered funding when I~applied
because it was thought that I~would just become a ``housewife'' after
graduation. I~knew that this (the part about my becoming a housewife) was
not true because I~had already tried that. Although I~believed the truth of
the stories, I~do not feel that it detracted from my overall positive
experience in the department. I~felt treated like an individual. That being
said, the atmosphere in the statistics department could be difficult at
times. It was a very small department and if faculty were feuding, it came
across to the students. We had a great department administrator, Louisa Van
Baalen, who was influential in keeping the atmosphere positive.

\textit{Ryan}: Tell us more about the people in the department in those
days? Were there any other women? What was the atmosphere like?

\textit{Laird}: When I~arrived in the Department, I~was one of four in the
entering Ph.D. class (the others were George Cobb, Diek Kryut and Sharon
Anderson). Persi Diaconis had entered the previous January, so he was
taking some courses with our cohort. The faculty were Bill Cochran, Fred
Mosteller, Art Dempster and Paul Holland. We were informed that since it
was Cochran's last year of teaching at Harvard, we had to take all his
courses straightaway. So we took Sample Surveys and Design of Experiments.
I~recall that the first day of class in Sample Surveys, Cochran stood at
the board and said (in a kind of indistinct mumble) that this was the most
boring topic in the world and he could not imagine why anyone would take
the class. I~also recall that I~went to his office to ask him questions
about the class. The first time was OK, but on the second occasion, he said
to me ``you are getting to be a bit of a nuisance.'' Having Cochran in the
department was really special and everyone treated him with the utmost
respect.

%f1 #&#
\begin{figure}%[b]

\includegraphics{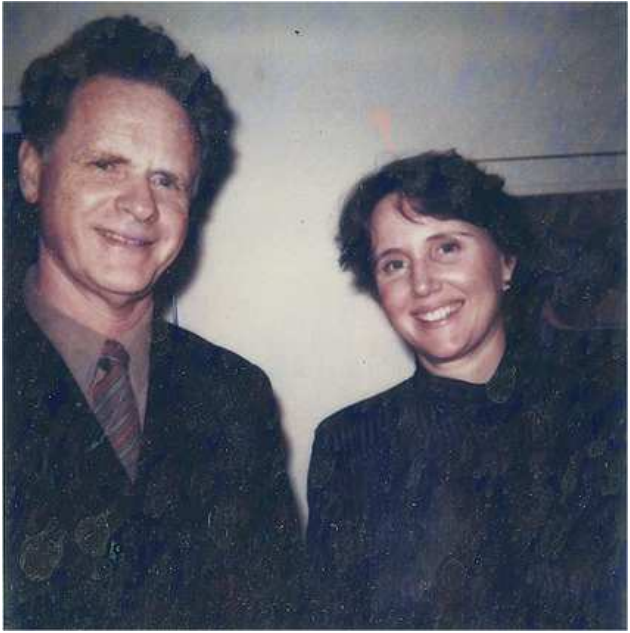}

\caption{Nan Laird with her Ph.D. advisor, Professor Arthur
Dempster, in the Harvard Statistics Department, circa 1975.}
\end{figure}

Paul Holland was another faculty member who was very involved in teaching
and engaged with students. I~really admired his teaching style.

Fred Mosteller was there and I~worked on his grant with Dave Hoaglin, as I~did not have any funding besides working as a teaching assistant. Fred was
involved in a lot of activities, national panels, teaching and
collaborating in other departments and schools. He had so many interests
and projects going on; he showed me that statistics was important
everywhere and statisticians could have a big impact in many fields.

\textit{Ryan}: That must have been a great experience working with Fred.
Tell us more.

\textit{Laird}: It was wonderful working with Fred. He treated all of his
graduate students like team members. You got to know about the projects he
was working on, and sometimes got to work with him on them. He was very
involved in a lot of activities and not so available as other faculty, yet
it was a privilege to work with him. Looking back on it, I~realize it was
like being a part of a statistics ``lab.''

One important lesson I~learned from Fred was about criticism. I~submitted
the first paper from my thesis to Biometrika. It came back asking for
changes. I~thought it was a negative rejection so I~just put it in a
drawer. After a while Fred asked me about it and I~said it was a negative
review. He asked to read the reviews, and on doing so said, ``This is not a
bad review, this is a good review.'' Of course, he was right; it was easy
to get the paper published after minor changes. Years later I~saw genuine
bad reviews and I~was grateful to Fred for teaching me the difference.

%f2 #&#
\begin{figure*}[t]

\includegraphics{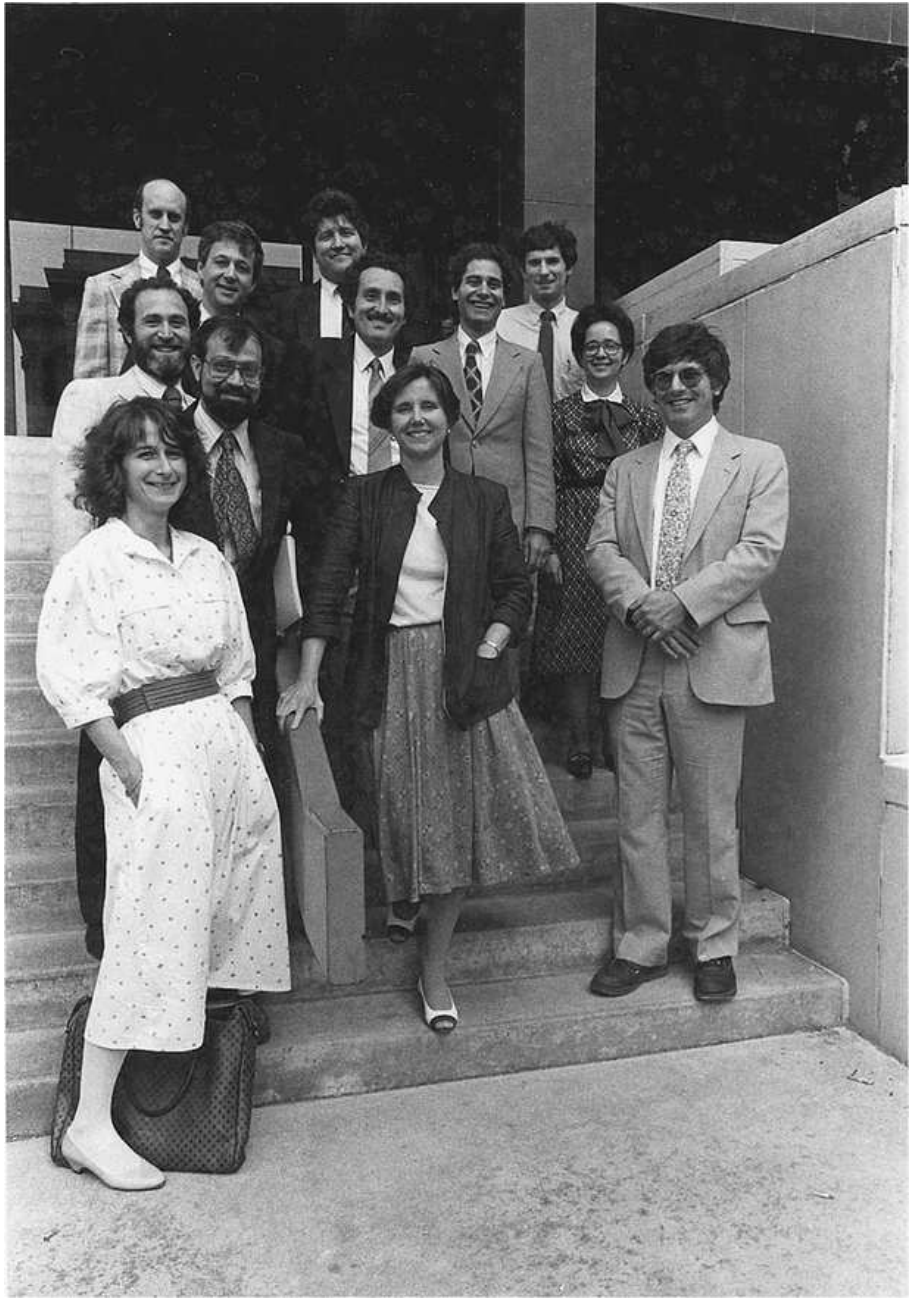}

\caption{Junior faculty members from the Biostatistics Department at the Harvard
School of Public Health, circa 1984. Back row: David Harrington, Stephen
Lagakos, Colin Begg; Second to Back row: Richard Gelber, David Schoenfeld,
Michael Feldstein, Thomas Louis, Rebecca Gelman; Front row: Christine
Waternaux, Cyrus Mehta, Nan Laird, Henry Feldman.}
\end{figure*}

%f3 #&#
\begin{figure*}

\includegraphics{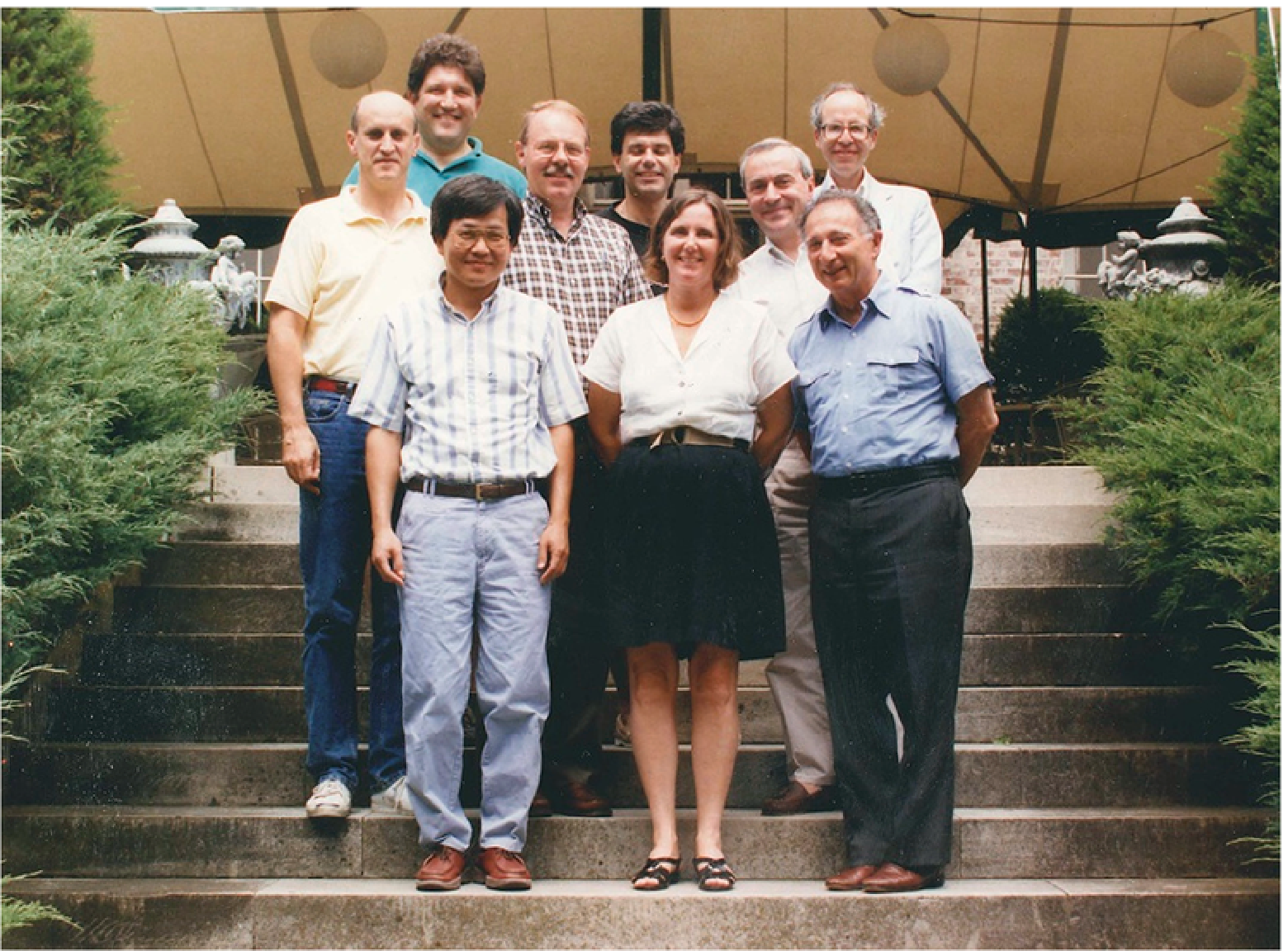}

\caption{Senior faculty members from the Biostatistics Department at the Harvard
School of Public Health when Nan took over as Chair in 1990. Back row:
David Harrington, Stephen Lagakos, James Ware, Butch Tsiatis, Marcello
Pagano, Bernard Rosner. Front row: L.~J.~Wei, Nan Laird, Marvin Zelen.}
\end{figure*}

\textit{Ryan}: Now I~know that Art Dempster was your Ph.D. supervisor. Tell
us about that.

\textit{Laird}: Art Dempster was the department chair at the time and was
basically the main ``go to'' advisor for all the students until you started
your thesis. For me, he was a great source of stability in the department.
He taught quite a few of our basic courses and I~found him accessible even
though some students did not. I~wanted to work with Art because I~liked the
way he thought and I~liked his approach to statistics. He was interested in
models and suggested that we work on models with random effects and
variance components. He was not a very ``hands on'' thesis advisor. I~recall that close to the end, within a few weeks of finishing my thesis, I~asked him a question about the last chapter. He responded by saying that he
really had not followed my work very closely so he did not think he could
help! I~got a big kick out of that.

\textit{Ryan}: How did the EM paper (\cite{1}) come
about? What was it like to work with Art Dempster and Don Rubin on that?
Did you have any idea of just how famous that paper would become?

\textit{Laird}: The idea of the EM algorithm came about during the last
year of my graduate studies. I~was working with Art on a random effects
approach to smooth two-way contingency tables. I~had a log-linear model for
the counts, then put a normal prior distribution on the log parameters with
zero mean and a single variance component. But I~was struggling with the
problem of how to estimate the variance component---now it seems trivial,
but at the time, computations were difficult and messy (we were still in
the punch card era). I~asked Art about estimating the variance component;
he said, ``Why don't you just use that substitution algorithm?'' I~had
never heard of the substitution algorithm, but after reading some papers
and further discussion with Art, I~went off and worked on the application
of this idea to my random effects problem.

I~discovered that the maximum likelihood estimate of the variance component
was a fixed point of the substitution algorithm, where the likelihood was
obtained by integrating out the random effects. So I~reported this back to
Art. After a characteristic short silence, Art said, ``Well, this
derivation will work for any exponential family likelihood with any kind of
missing data.'' That was the origin of the EM for me. At the time Art was
also working on missing data with Don Rubin, who was at the Educational
Testing Service (ETS). Don received his Ph.D. from Harvard a couple of years
before I~came, so I~did not know him. Art suggested that the three of us
write a paper together. Art and I~drove to Princeton early on to work with
Don. That was the first time I~recall meeting Don.

I~did realize that the EM was a very important contribution, and likely to
be a famous paper. But the EM paper was to be one of my first real papers
and I~didn't want to trade on that my whole life, so I~was pretty intent on
branching out into other areas as well.

\textit{Ryan}: I~remember from when I~started there in 1979 that you were
still somewhat involved in the statistics department then, but gradually
your activities all moved over to the Harvard School of Public Health
(HSPH). How did that come about?

\textit{Laird}: I~completed my Ph.D. in 1975 and took a position as an
assistant professor in the Biostatistics Department at HSPH. Initially, I~was secondary with statistics so I~taught and advised students there for a
few years, but eventually found my work at HSPH was very engaging.

\textit{Ryan}: You were there during a time of major change and transition,
not just for the department, but I~think for the field of biostatistics as
well. You would have been there when Fred Mosteller became chair and in
those famous (or should I~say infamous?) days when Marvin Zelen arrived en
masse from Buffalo. What was the Department like in those early days?

\textit{Laird}: My first few years in the Biostatistics Department were
interesting. There actually were quite a few women there when I~came---Jane
Worcester was chair and both Marge Drolette and Yvonne Bishop were on the
faculty. Bob Reed was the other senior faculty person. I~think most people
like to do something that matters to the lives of others, but perhaps women
more so and this is why we are drawn to application areas such as health.
The department was pretty sleepy in those days, with the faculty having
heavy teaching loads or involvement in collaborative projects. When I~joined, Jane Worcester was very close to retirement and there was a lot of
speculation about who would be the next chair. Howard Hiatt was a new Dean
then, and he was interested in strengthening Biostatistics. Much to my
surprise, Fred Mosteller took on the job for a few short years, and really
did change things. One thing he did was to give offices to all of the
Biostatistics faculty. Another was to hire a lot of junior faculty---Jim
Ware, Larry Thibodeau, Tom Louis, Christine Waternaux were some that I~recall. This was great for me because I~had colleagues who were
contemporary in age and had similar concerns and career goals. Of course,
the really big recruitment was Marvin Zelen and his team from Buffalo.

\textit{Ryan}: Tell us more about that. I~understand that he brought along
quite a few people with him, including a number of young up-and-coming
statisticians such as Steve Lagakos, Richard Gelber, Colin Begg and Rebecca
Gelman, to name a few. In an interview for the Department newsletter in
1997, you said that some referred to ``Marvin's Baseball Team.'' Marvin's
arrival was also the start of a strong connection between the department
and the Dana-Farber Cancer Institute, which still exists today. I~imagine
this must have been a time of rather mixed feelings---excitement at all
the new possibilities, along with the natural anxiety about change. How did
the department and the school as a whole react? How did the department
change as a result of these developments?

\textit{Laird}: Well, there were many discussions before he came, both in
the Department and within the School. Yes, there was anxiety about their
coming. Faculty outside the department were worried that Biostatistics
would ``take over the school.'' Faculty inside the department were worried
about their futures and that the reward system might be different. But
Marvin's group came, and they assimilated. Some people left, but there were
losses on both sides, and eventually gains on each side. Marvin took over
as department chair a few years later. I~thought he was a great chair
because he made it clear that the academic program belonged to the faculty
and he expected all of us to participate. He set very concrete achievable
goals for the department, like strengthening the doctoral program and
obtaining training grants. These were goals that everyone could unite
behind. He also was very principled, especially about salary equity. I~received big raises after Marvin took over as chair, because my salary was
far below that of my colleagues at the same rank.

%s3 #&#
\section{Longitudinal}
\label{sec3}

\textit{Ryan}: I'm interested in your comment about most people liking to
work on something that matters. For many of us, that translates to working
on applications. But I~have always admired your ability to take a slightly
more abstract approach and come up with tools and methods that have broad
application. I~think it takes confidence to believe that one's methods will
matter (as opposed to solving a specific real world problem). The EM is a
great example, but so is your work in the early 80s with Jim Ware. In
particular, your growth curve paper (\cite{6}) has had major
impact. Tell us more about that.

\textit{Laird}: Yes, I~have always been interested in looking for a general
framework, rather than working on a series of smaller ``one-off'' problems.
In a way, working on some theory is easier because you can say how the
method will behave under specific conditions. With applications involving
data, so often the data refuses to cooperate. The answers you get can
depend more on characteristics of the data and not the statistical method.
I've always liked developing methods that others can apply. In the case of
the growth curve modeling, I~had started working with Jim because of his
interest in applying random effects models in a longitudinal cohort called
the ``Six Cities Study.''\footnote{
This was a prospective cohort mortality study involving 8111 randomly
selected residents of six U.S. cities. The objective was to estimate the
effects of air pollution on mortality and lung function while controlling
for other risk factors such as the individuals' smoking status and age
(\cite{4}).} These were the days when longitudinal data analysis was just starting to
become popular. The Six Cities Study had a lot of unbalanced data and some
missing data as well, whereas all the textbooks were providing solutions
for balanced complete data. Using the random effect structure was a natural
and it tied in well with the EM algorithm.

Jim and I~worked on lots of interesting problems with colleagues such as
Tom Louis and Christine Waternaux, and we also co-advised several doctoral
students (Fong Wang, Nick Lange, Christl Donnelly, Rob Stiratelli, Masahiro
Takeuchi) on longitudinal methods. I~think one reason our 1982 paper was so
popular was because Jim had a really good fix on issues that applied
statisticians were grappling with. Jim has always had a great sense of
where a field is headed and has an excellent eye for good problems. Christl
Donnelly had an interesting thesis, she applied the mixed model to estimate
the spatial distribution of air pollution, using an approach closely
related to Kriging.

%f4 #&#
\begin{figure}

\includegraphics{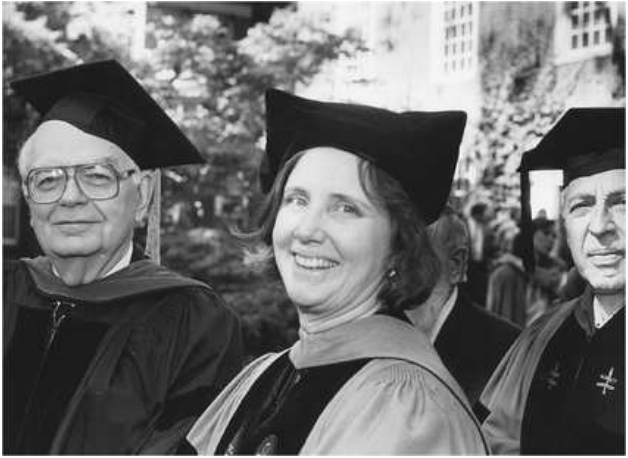}

\caption{Nan Laird with Fred Mosteller (left) on the awarding of his Honorary
Doctorate at Harvard, 1991.}
\end{figure}

This collaboration with Jim was really important for me. I~learned about
applying for grants from the National Institutes of Health (NIH). No one
had ever mentioned grant support to me before, let alone encouraged me to
get my own. Jim had been at the NIH before coming to HSPH, so he had good
idea about how things worked. I~learned so much by working with Jim. Our
families also became close and we spent a lot of time together outside of
work.

\textit{Ryan}: I~recall reading the Laird and Ware paper many years ago and
struggling to understand the way you were applying the EM. In fact, it was
only after attending Xiao-Li Meng's thesis defense that I~realized that you
were actually using a technique that hadn't been invented yet, namely, the
ECM (\cite{11}).

\textit{Laird}: Yes, when we applied the EM strictly in the mixed model
regression context, estimating the regression and variance parameters
together, the algorithm did not take advantage of the well-known closed
form for the regression parameters, which is available if the variance
parameters are known. We did something that seemed really natural, namely,
estimating the regression parameters through weighted least squares,
\mbox{assuming} the variance parameters were known, then estimating the variance
parameters via an EM, but assuming the regression coefficients were known.
This modification worked well, and was, as you say, an example of the yet
to be invented ECM.

\textit{Ryan}: You continued working on longitudinal data analysis for many
years.

\textit{Laird}: In addition to the work on repeated measures that Jim and I~did, I~was also interested in models for repeated categorical data. I~had
several students who worked in this area. Stuart Lipsitz and Garrett
Fitzmaurice both worked on parameterizations for repeated categorical
outcomes, and the interplay between the mean effects and the covariance
parameters, using the so-called marginal models. Jarek Harezlak was one of
my students who very independently found his own topic on applied
functional analysis in the longitudinal data setting. I~think I~learned
more from him than he did from me! Skip Olsen looked at adjusting for the
baseline measure when the interest is change over time.

%s4 #&#
\section{Meta-Analysis}
\label{sec4}

\textit{Ryan}: I~enjoyed reading your account in the recent COPSS book
(\cite{9}) of writing your papers with Rebecca DerSimonian on
meta-analysis (\citeauthor{2}, \citeyear{2}, \citeyear{3}). Tell us a bit more about
this.

\textit{Laird}: Yes, the paper with Rebecca in \textit{Clinical Trials} is
one of my most cited. That has been a big surprise for me, in part because
it is such a sleeper. The citations are now growing exponentially, but it
came about something by accident. Fred Mosteller asked me to have a look at
a paper (\cite{12}) that combined a series of 23 studies on
the impact of coaching to improve SAT (Scholastic Aptitude Test) scores.
They concluded that coaching helped, thereby contradicting the principle
that the SAT measured innate ability, but were advised to consult with a
biostatistician. I~read the paper and was struck by the amount of
variability in the studies as well as how much the results varied by study.
There was also variation in study design: some studies had no control
group, some were observational, a few were randomized and a few involved
matching. The degree of control in the coached group seemed to be inversely
related to the magnitude of effect. For example, some studies only
evaluated coached students, comparing their scores before and after
coaching. So all this got me thinking that it would be useful to develop a
formal statistical framework for\vadjust{\goodbreak} the analysis. Rebecca Dersimonian was my
graduate student at the time, working on random effects modeling. I~got her
working on a modeling framework for meta-analysis that allowed for
study-to-study variation in effect sizes through the inclusion of random
effects. We reanalyzed Slack and Porter's data, concluding that any effects
of coaching were too small to be of practical importance. While our paper
in the \textit{Harvard Education Review} never got that many citations in
the scientific literature, it got a lot media attention.

A few years later, Rebecca and I~published a follow-up paper in
\textit{Controlled Clinical Trials}, applying the method to the clinical
trial setting and this paper has been very highly cited. It is sometimes
cited as ``the standard'' approach to meta-analysis. A nice feature of our
approach was that it yielded a closed-form estimator and I~think this is
one of the reasons why the paper got so much interest. Another feature is
that it shows how to do the analysis just by using data summaries.

\textit{Ryan}: Back in the 90s, I~remember talking with you about a study I~was doing with Lew Holmes from Massachusetts General Hospital on a
meta-analysis of adverse effects associated with chorionic villus sampling
(CVS). We wanted your advice on the wisdom of going after individual level
data versus working with summary statistics based on published papers.

\textit{Laird}: Well, I~don't really remember the details of that. Danyu
Lin (\cite{10}) says that it shouldn't really matter---if the
models are the same, then the results should be the same whether you use
individual level data or data summaries. But I~find researchers often
overemphasize the importance of individual level data. I~think that is a
``case study'' rather than a statistical perspective. In practice, it is
probably more important to focus on what study characteristics help to
explain study-to-study variation in the observed effects. In addition, it
can be very hard to get individual level data.

\textit{Ryan}: Yes, I~agree! In the case of our CVS study, we did manage to
get individual level data for a few studies, but they tended to be the
studies where nothing at all interesting was happening$\dots.$ Perhaps an
inverse of publication bias? What are your thoughts about trying to take
account of study quality in the context of meta-analysis?

%f5 #&#
\begin{figure*}

\includegraphics{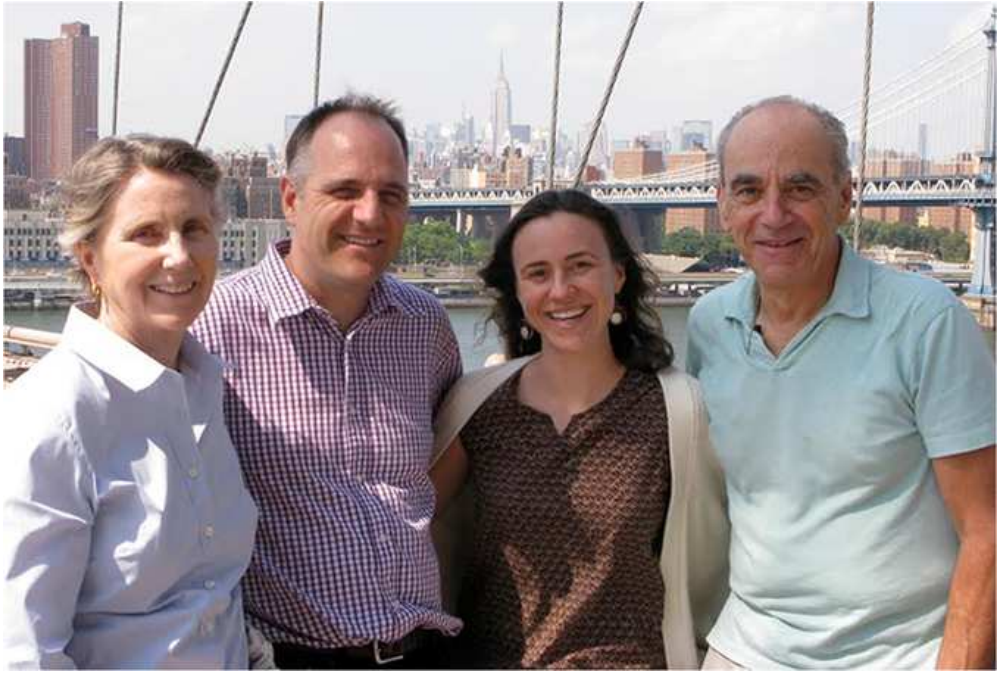}

\caption{Nan Laird with her family in 2009. From left to right, Nan, Richard Hughes,
Lily Altstein and Joel Altstein. Lily is a biostatistician at Massachusetts
General Hospital and married to Cory Zigler who is an assistant professor
of Biostatistics at the Harvard T. H. Chan School of Public Health. Richard
is a chef in Concord, Massachusetts, and Joel is a Real Estate Developer in
Cambridge, Massachusetts.}
\end{figure*}

\textit{Laird}: In my paper with Rebecca, we originally stratified by study
design and then looked at the distribution of effect sizes. Study design
generally explains a lot.

%s5 #&#
\section{Psychiatric-Epidemiology}\label{sec5}

\textit{Ryan}: Let's talk about some of your other work from the 1980s and
1990s. I~know you continued your work on missing data, but you also got very
interested in methods for application in psychiatric epidemiology. How did
that come about?

\textit{Laird}: Christine Waternaux moved her main appointment to McLean
Hospital, one of Harvard's teaching hospitals specializing in psychiatric
disorders. Christine was interested in submitting a training grant
application to the National Institutes of Mental Health. The Department of
Epidemiology had one for a long time, but it had lapsed. Christine and I~collaborated with the epidemiology department, successfully bringing in a
new training grant from the National Institute of Mental Health (NIMH) that
supported both epidemiology and biostatistics students and postdoctoral
fellows.

\textit{Ryan}: Tell us more about how the grant worked. It can be a
challenge to have a grant that crosses two departments like that. What were
some of the interesting problems that arose for you and your students?

\textit{Laird}: The grant gave us the resources and the contacts to get
involved in a number of exciting projects. What to do with multiple
informants comes up a lot in the psychiatric setting, where assessments of
study participants may be provided by the individual themselves, their
care-giver or a relative. Garrett Fitzmaurice and Nick Horton both had a
background in psychology which made their work on the NIMH training grant
very valuable. Garrett's thesis work on repeated categorical outcomes
extended naturally to this multiple informant setting. The interesting fact
about his approach was that it could apply to either the outcome or to the
covariate, or both, and this comes in handy in the multiple informants
setting as well. Nick Horton also worked on multiple informants and the
three of us, plus Stuart Lipsitz and Sharon-Lise Normand, developed a nice
set of tools for handling multiple informant problems, especially for
applications in psychiatry and health services research.

My background with methods for longitudinal data led to an involvement in
the Stirling County Study looking at depression and anxiety, with data
collected in three cross-sectional waves as well as longitudinally. Nick
Horton and Heather Litman both worked on Stirling County Study data as a
part of their doctorates. Kristin Javaras and Jim Hudson were both great
postdocs who were later supported by the NIMH training grant. The
psychiatric-epidemiology work led naturally to lots of interesting problems
with missing data, especially drop out. Often, the dropout was thought to
be informative.

%s6 #&#
\section{Missing Data}

\textit{Ryan}: Yes, tell us more about that. Given your background with the
EM, it seems like a natural area for you.

\textit{Laird}: Yes, missing data is a big issue in longitudinal studies,
but happens just as well in cross-sectional studies. Certainly the EM
algorithm is a natural way to handle missing data in lots of settings, but
it can be problematic when missingness is informative. Many people at that
time were interested in the so-called selection model approach, which
required specifying the probability of dropout, conditional on both
observed and unobserved data. The approach is very appealing because it
allows you to specify familiar models, but it is highly parametric and it
is not really possible to test it empirically, nor to understand the
limitations of the model. Stuart Baker and I~worked on a selection model
for categorical data in the logistic regression framework, and obtained
some interesting results about model identifiability. Bob Glynn also worked
on nonignorable nonresponse in the sample survey setting. Joe Hogan worked
on nonignorable dropouts in longitudinal studies, but using a completely
different modeling strategy, one related to the pattern-mixture models. The
idea was to reverse the conditioning to look at the distribution of
observed data given each particular pattern of missingness, and then to
integrate over the various patterns to get the marginal distribution of
interest. With this approach it is easier to expose the model
assumptions.

%s7 #&#
\section{Becoming Chair}\label{sec7}

\textit{Ryan}: You became Chair of the Harvard Biostatistics Department in
1990. In a tribute to your accomplishments during your nine years as chair,
Jim Ware noted how much the department grew under your leadership. A lot of
innovations happened in that nine year period: the Center for Biostatistics
in AIDS Research (CBAR) was established under Steve Lagakos' leadership,
the Department's minority training program got started. Tell us more about
those years, in particular, what you feel were your biggest
accomplishments.

\textit{Laird}: I~was excited to be department chair. Of course, Marvin
Zelen was a very difficult act to follow, but I~had a very different
personality and different ways of doing things, so I~never felt that I~was
trying to replicate what he did. Marvin had a lot of little ``tricks'' that
were part of what made him such an effective chair, and I~used a lot of
those myself. One thing he did was turn over the running of the department
to the faculty by setting up a series of committees that did all the
department work. We had the degree program committee that set policy about
requirements for degrees, the curriculum committee that decided what
courses were taught, the student advising committee that set policy about
students, like stipends, and assigned advisors. This meant that Marvin did
not have to make all the day to day decisions about operations. But more
importantly, it gave the faculty power over running the department. Marvin
rarely interfered with department business, but if there was an issue he
cared about, he let the faculty vote; he just set up his votes in advance
with lobbying so everyone knew what he wanted! I~thought these were great
strategies and did my best to apply them as well. My goals were broadening
our research agenda and strengthening faculty hires, especially since we
were expanding at a rapid rate with the initiation of CBAR and the AIDS
work. I~felt we needed to improve the environment for new faculty. One
thing I~initiated was the policy that every new junior faculty hire have
support for at least 40\% of their time in their first 3 years to develop
their own research agenda. Over the years the 40\% has fluctuated, but the
principle remains.

I~also increased the Department's activity in statistical genetics. This
was about the time of the completion of the Human Genome Project and the
field was highly visible and changing rapidly. As department chair I~brought in a lot of outside speakers in the area, obtained grant support
for genetic methods and developed several important collaborations.

Initiating the 40\% research policy for junior faculty that allowed them to
strengthen their own research agenda, and developing statistical genetics
as a research area for the department were my two main contributions as
department chair.

%s8 #&#
\section{Statistical Genetics}
\label{sec8}

\textit{Ryan}: You also got very involved personally in statistical
genetics research. Tell me about that, in particular, what motivated that
move for you?

\textit{Laird}: I~was intrigued by genetics ever since writing the EM paper
because many of the earliest applications of EM were in genetics. It is
only natural because, until relatively recently, one could not directly
observe DNA. I~had a sabbatical at the end of my first five years as
department chair and decided to spend it learning more about problems in
genetics; this proved to be very fruitful. One problem that caught my eye
was using family data to test whether or not a particular genetic locus was
involved in causing disease. At that time, people were using the TDT
(Transition Disequilibrium Test) that required genetic data on parents and
their diseased child. But the question arose as to how to generalize this
to using siblings and other family members when parents were not available.
This was common in late onset disease such as Alzheimer's. I~collaborated
with several talented students, postdoctoral fellows and faculty to develop
a general framework for family designs and developed the FBAT software
which has been quite popular. Steve Horvath, Steve Lake and Christoph Lange
extended the theory and the program, adding a lot of regression ideas for
measured disease outcomes. Tom Hoffmann extended the work that Steve Lake
did on gene--environment interaction studies. I~had quite a few other
students over the years working in genetics: Xiaolin Wang, who was my first
student in this area, Ronnie Sebro, who also worked with Neil Risch, Cyril
Rakovski, Xiao Ding, Gourab De. Currently, Christina McIntosh is looking at
design issues in family studies with multiple affected individuals.

Christoph Lange and I~were actually involved in one of the earliest
genome-wide studies. We were looking at the family component of the
Framingham Study with a view to identifying genes associated with obesity.
We had data from a chip with 100,000 SNPs (single neucleotide
polymorphisms). Christoph had an idea for a method that exploited the
independence of within and between family information. We used the between
family information to estimate the power of each SNP test, then chose only
the top ten for testing. This enabled us to avoid the loss of power
associated with the multiple testing problem. We found a SNP near the
\mbox{INSIG2} gene associated with obesity. The results were controversial and
many investigators attempted replications. Some replicated our results,
while others did not. A~follow-up meta-analysis showed that the association
is present in studies of general populations, but that the reverse
association is seen in studies involving populations selected to be
healthy.

I~had a somewhat similar experience when I~was involved in a study that
identified a potential Alzheimer's gene. There were a number of follow-up
studies, some that replicated, some that didn't. Genetic data are tricky.
It is easy to find lots of associated SNPs, but really hard to pin down
causal mechanisms in the DNA sequence. With complex diseases it is very
difficult to attribute major effects to one small disruption in the DNA. A
few dramatic cases (Mendelian disorders) have influenced people to think
they will find a variation in the DNA coding that explains what's going on.
In reality, it's most likely a complex interplay between multiple genetic
variants and environmental factors.

There is a lot of variation in the quality of papers appearing in the
genetics and statistical genetics literature and, truth be told, there's
not a lot of quality control. People are under a lot of pressure to
publish, and referees are under pressure to do fast reviews, but there are
so many pitfalls. Genes are very complex---you can look at gene
expression, DNA, exons, introns, methylation sites etc.---and there is huge
potential for false positives. The problem is exacerbated by the fact that
environmental exposures are notoriously hard to measure accurately, even
when we know what the relevant ones are. I~think it is important to build
genetic investigations around a theory, as we commonly do in
epidemiology.

%f6 #&#
\begin{figure}

\includegraphics{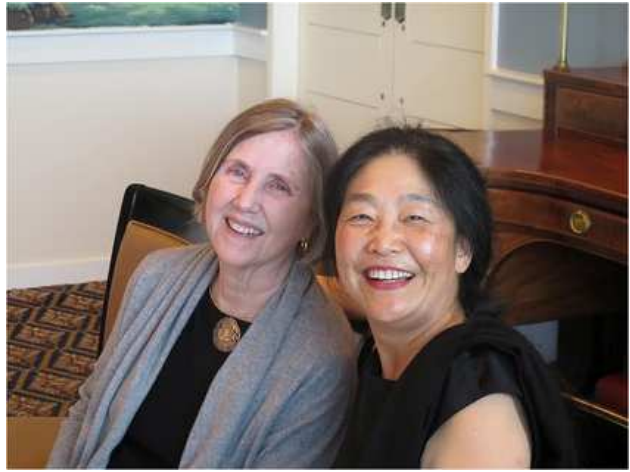}

\caption{Fong Wang-Clow with Nan at her retirement celebration, May 2015.}
\end{figure}

\textit{Ryan}: It seems these days that because of technologies like GWAS,
people don't want to pin down their hypotheses in advance any more.

\textit{Laird}: I~agree---the general null hypothesis of no genetic
variant affects the outcome is not so useful. There are always adjustments
such as Bonferroni or FDR approaches, split sample etc., but power is often
low. People are also now starting to impute SNPs in an effort to boost
power by increasing coverage. I~think these issues need a lot more
statistical attention.

But I~think a bigger problem is that people don't take account of design.
People have done a lot of convenience sampling, obtaining biological
samples from preexisting studies designed for different purposes.

%s9 #&#
\section{What's Next Workwise?}
\label{sec9}

\textit{Laird}: Looking back, I~think that venturing into the genetics area
took over my career. But I~enjoyed working in the area. It is exciting when
you can use knowledge of how the biology of inheritance works to drive the
analysis. My other work hasn't had those strong biological underpinnings.

\textit{Ryan}: So do you think you'll keep working in genetics?

\textit{Laird}: I'm looking forward to working on whatever I~like. That may
or may not include genetics.

\textit{Ryan}: Yes, I~had a couple more questions about that. Random
effects modeling features very prominently in your work. Were you ever
tempted to become a Bayesian?

\textit{Laird}: No, I~was not. The business of the prior always seemed to
get in the way.

\textit{Ryan}: Even working with Art Dempster?

\textit{Laird}: Is Art a Bayesian? Certainly when I~worked with him I~would
not have called him strictly Bayes. I~would say he emphasized the
likelihood, but if he used priors, it was usually flat prior and often
empirical Bayes. My early work was always about likelihood-based methods. I~suppose a lot of my work could be described as empirical Bayes because we
were dealing with hierarchical models. But we never specified hyperpriors,
just used the data to estimate unknown parameters such as variance
components. My work in genetics, however, has been largely score-based
tests. In this setting we could obtain the distribution of the data under
H0 using Mendel's laws, so it was a natural approach.

\textit{Ryan}: What about survival analysis?

\textit{Laird}: Yes, I~did one paper with Don Olivier (\cite{8}). We formulated the problem as piecewise exponential and made a link
to log-linear models. It was a fun paper and provided a way to fit models
to survival data in the days when general survival software wasn't
available. It also provided a simple way to test for nonproportional
hazards.

\textit{Ryan}: Yes, that paper was very useful. I~remember reading it and
using some of the ideas with one of my students who was working on the
analysis of carcinogenicity data. One of the many things I~have admired
about you over the years Nan is that you don't ``tweak''---your papers
all have real impact.

\textit{Laird}: Tweaking? Oh, I've tweaked, Louise! Whenever I~felt that I~was starting to tweak, that's when I~moved to a new area. It's part of the
reason why I~think it's time for me to start to wind down, research wise.
But really, I'd like to explore other ways to stay engaged
professionally.

\textit{Ryan}: Tell us more.

\textit{Laird}: Well I've always really enjoyed teaching and working with
students. I~like collaborating as well. I~have also greatly enjoyed my
collaborations. I~will continue to work with colleagues at the Channing
Labs (Harvard Medical School) where we are looking at the genetics of
Chronic Obstructive Lung Disease.

In the early 1980s I~started collaborating with colleagues at Harvard's
SPH, Medical School, School of Government and the Law School. New York
State wanted an empirical basis for establishing reform of their laws on
medical malpractice, and we undertook a sample survey of hospitalizations
in New York to determine the rate of adverse events, the percent due to
negligence, and the percentage of adverse events that are never reported.
It was a landmark study and the first rigorous one of its kind. It still
serves as a model for tort reform. Russ Localio, a former student now at U.~Penn, and I~were the statisticians involved.

I~would like to get involved again with some National Academy or Institute
of Medicine panels. One panel I~worked with in 1986 was The Airliner Cabin
Environment, headed by Tom Chalmers. I~remember this as one of the most fun
and rewarding panels I~was on. The senate had asked for an investigation of
the quality of air on planes. It was the era of smoking sections in the
back, and, due to economy measures, airlines were starting to use
recirculated air in the cabins. I~recall thinking right at the beginning
that the simplest solution was to eliminate smoking, but it took slogging
through a lot of data to get to that recommendation. I~recall Dimitri
Trichopoulos (then future chair of HSPH Epidemiology Department) talking
about his work on the very small but important risks of side stream
cigarette smoke, and Brian MacMahon's (then current chair of HSPH
Epidemiology Department) belief that any odds ratio less than 2 was too
small to be of any relevance. In the end, all of us on the committee were
convinced that a smoking ban was in order. The senate was also convinced,
and it was enacted into law shortly thereafter. This was certainly one of
my most signal public health contributions. Although we did not do a formal
meta-analysis of the health effects of side-stream smoke, this experience
influenced both me and Tom Chalmers about the importance of meta-analysis
for public health.

%s10 #&#
\section{Books}\label{sec10}

\textit{Ryan}: You've co-authored several already, including one on
longitudinal data (\cite{5}) and one on
statistical genetics (\cite{7}). Are there likely to be
more?

\textit{Laird}: I~am very proud of the books and enjoyed those
collaborations a great deal, but writing a book is very demanding. There is
also an IMS Monograph on Longitudinal and clustered data. The book with
Christoph is about statistical genetics, but written from the perspective
of a statistician. It lays out some of the fundamental ideas that drove
statistical genetics, starting with linkage, then moving into family-based
designs and GWAS, but stopping short of sequence analysis. It's written for
the graduate level, but is fairly applied and has been used by a lot of
people for teaching.

\textit{Ryan}: The book on longitudinal is more classically statistical,
right? How did that one come about?

\textit{Laird}: After working with Jim for a few years, both on research
and on teaching short courses mostly outside of Harvard, I~decided to teach
a regular course for the graduate students here in the department on
longitudinal data analysis. The course was rigorous, but pitched at a level
accessible to both biostatistics and epidemiology students. After I'd
taught it a couple of times, Jim and then Garrett took over teaching the
course. The book evolved from the course notes and it is very popular. I~think Garrett deserves a lot of credit for the book's success. He is a
wonderful co-author. He's a very good writer and put in a huge amount of
work on the book, building in lots of detailed examples as well as SAS and
R code. He always took Jim's and my input constructively. The book is much
broader than growth curve modeling, including GEEs as well as clustered
data and multiple informants. The book is very pedagogical and
comprehensive. It captures a lot of my professional life and I~suppose
that's another reason why I~feel very proud of the book.

\textit{Ryan}: So many of your students have done spectacularly well. You
must feel proud. I~recall a speech that Garrett gave a few years back at a
dinner in your honor, where he talked about what a fantastic mentor you
are. What do you think it takes to be a good mentor? Did you have good
mentors? What are some of your secrets?

\textit{Laird}: Yes, my students have done beautifully, some spectacularly,
and I~am very proud of them all. I~suppose I~always try to be straight with
my students. When they are first starting out, I~try to give them a lot of
help if they need it, but gradually I~step back, letting them take more of
a lead. I~think that's important if they are to become independent
researchers eventually. I~always tried to put aside an hour a week to talk
with students and I~like to make them feel connected to the research area.
So I~stress reading papers not only clearly related, but peripherally
related also, or going to seminars.

Mentoring students has always been one of the favorite parts of my work.
This is one reason that I~chose to have my recent retirement celebration
focused entirely on my students. It was organized by Garrett Fitzmaurice
and Christoph Lange, both of whom spoke, as did Rebecca DerSimonian, Joe
Hogan and Kristin Javaras. The highlight was Fong Wang-Clow, who with her
husband Eric, donated \$1,000,000 to the department to establish the Fong
Clow Doctoral Fellowship in my honor. I~was very surprised and honored! The
Biostatistics and Epidemiology Psychiatric Seminar Group also held a
special seminar honoring my retirement and Steve Lake spoke there.

In terms of my own mentors, I've certainly been influenced by some powerful
senior statisticians, such as Fred Mosteller, Art Dempster and Marvin
Zelen. But other than the EM with Art, and one meta-analysis review article
with Fred, they were not collaborators. I~think I~learned different sorts
of things from each one of them. Strictly speaking, Jim Ware was a peer
rather than a mentor, but I~learned a lot from working closely with him
over the years.

While of course neither of them was my student, both my daughter Lily and
her husband Cory Zigler have their Ph.D.'s in Biostatistics (from UCLA).
While I~don't feel I~can take much credit for their professional
development, I~am very proud and happy to have two other biostatisticians
in the family!

%s11 #&#
\section{Balancing Work and Family Life}
\label{sec11}

\textit{Ryan}: Many of your colleagues, for example, Jim Ware, Tom Louis
and Butch Tsiatis, were quite active in the professional societies, for
example, serving as President of ENAR or the ASA. You never seemed to do
much of that. How come?

\textit{Laird}: I~had to make choices early in my career. My first husband
and I~separated when I~entered graduate school, and, as a single mother, I~jealously guarded my time for the sake of my son. This feeling stayed with
me throughout my career and after I~remarried. I~felt my family deserved my
attention and I~tried to be a 9 to 5 person. I~was not very successful with
this, especially when I~was chair, but I~thought it was important to try.

\textit{Ryan}: I'm impressed and even a little surprised since I've never
heard you express those thoughts, even after all these years that we've
known each other! So do you think it is possible to balance the demands of
career and family?

\textit{Laird}: I~have always been a bit private about personal things at
work. I~felt that I~came to the office to work, not socialize. It was
always a surprise to me that many men do socialize at work. If you have
small children, you have a big responsibility and I~felt it was important
to clarify the boundaries between work and family. Looking around at my
colleagues over the years, most of them were out of town a lot, sometimes
working, but often giving seminars and presenting at conferences. Through
Fred Mosteller, I~was involved in several National Research Council Panels.
This takes a lot of time in terms of travel and preparation. I~had to make
hard choices, and, understandably, I~decided to eliminate activities that
did not lead to academic publications or teaching success. All it takes is
saying ``No'' a few times, but I~do regret that I~wasn't able to get more
involved in pro-bono statistics work $\dots.$ I'd like to do more of that
now.

I~understand that now, in order to be an elected fellow of the ASA, you
need to have a track record of service to the ASA---I~would never qualify
now, but I~do feel I~have made contributions to our profession.

\textit{Ryan}: Well, of course, it helps when you are really smart and when
your work is so outstanding that it still gets recognized, even without
your doing the kinds of promotional things (talks, conferences etc.) that
most of us need to do to get our names out there! But seriously, I~think
your story is a touching one and points to the real challenges that face
women who have family responsibilities, yet want to pursue a career as
well. Attending conferences and giving talks are good examples of some of
the traditional markers of success that are very ``male.''

\textit{Laird}: Yes, I~agree. I~find it hard to be comfortable with
self-promotion, but sometimes you just have to do it. It can be worse to be
silent when you are passed over. I~remember early on in my career I~had
applied for a grant from the National Science Foundation. Nancy Flournoy
called me up to say that I~hadn't suggested any suitable reviewers. It
didn't even occur to me that one could do that! So I~really appreciated
Nancy's call. And I~really admire how forthright Nancy is.

I~think also that there are different phases of your career. Careers are
long, and last much longer than a childhood. You need to pick and choose
what you want and when you want it. When I~tell this to graduate students
today, they can be horrified that you might need to cut back on certain
activities for 10--15 years if you want to have kids. That seems impossibly
long for a young person starting out in their career, but of course, it is
really short. But I~think it does point to the difficulty with the tenure
clock coinciding with the biological one.

I~also think the situation is changing a lot for men as well. They are
expected to be equal partners in parenting, and this will change their
attitude toward careers. I~was lucky to have a husband who took on a very
large share of parenting that allowed me to pursue a career. And he was
proud of my career. I~think this is more common now than it was forty years
ago.

\textit{Ryan}: Did you try to connect with other women scientists over the
years? Was this something important for you?

\textit{Laird}: Yes, but it just happened naturally without my trying hard
to make it happen. I~was good friends with several female faculty members
here at HSPH. There are times you need to talk with another woman, though
there weren't that many around, and certainly not many with children. It
was pretty difficult, especially as a single mother when I~was in grad
school.

\textit{Ryan}: Yes, I~can't begin to imagine. Your story will be inspiring
for a lot of young women starting out in their careers. I~think your advice
about just saying ``No'' is really powerful. I~need to remember that
myself! I~had planned to ask you why you had stayed at Harvard all these
years instead of moving around like so many people do. Was that also
influenced by your family circumstances?

\textit{Laird}: Yes, a lot of the reason for staying at Harvard was
personal. I~never felt it was fair to disrupt the family. On the other
hand, Harvard is a pretty amazing environment. I~always felt a lot of
freedom to explore new ideas. Plus, my family has always been a
tremendously positive source of support in my career. When I~got tenure and
was interviewed for the HSPH newsletter, it was really important to me to
acknowledge the role of my family. The editor questioned putting this in
the newsletter, I~think because men did not usually do this sort of thing,
but I~insisted. I~remember being struck by the comments from a young woman
faculty member years ago at a panel discussion on the challenges of
balancing career and family. This young woman said that to her, it wasn't a
challenge, but rather a real plus since her family gave her the support and
encouragement she needed to succeed. I've always felt that way too. It
probably helped, though, that Joel was in a different field. I~know there
are lots of couples who are in similar fields and that can complicate
things.

%s12 #&#
\section{Wrapping up}
\label{sec12}
\textit{Ryan}: So, let's move toward wrapping up now. Tell us about some
more of the things you like to do outside of work.

\textit{Laird}: I~love being with friends and family, especially my two
granddaughters, Margaret and Gene\-vieve. I~look forward to spending more
time with them and watching them grow. It is amazing how each child is so
different. I~can see why studies on the effect of environmental influences
on children's development are so hard to do.

I~am a passionate gardener and I~have thought about learning more about
landscape design. I~want to travel and see some places for pleasure rather
than just for work I~like to sew, especially quilts. Quilt design is very
appealing to me because it is so geometric and colorful.

\textit{Ryan}: Let's finish up spending just a bit of time talking about
where you see our field heading. A lot seems to be happening these days and
I've heard many people say that statisticians are getting left behind by
data miners, data scientists and the like. What are your thoughts?

\textit{Laird}: One of the things I~see, and this is especially so for
statistical genetics and bioinformatics, is that people from fields like
physics and computer science are taking over the computations.
Statisticians can sometimes be hampered by their desire to get methods
exactly right in terms of their properties. In contrast, people in other
fields are happy to get an approximate answer. We statisticians have such a
tradition of basing our publications on strong theory. I~am impressed with
the rigor, but the result is that much of our work is not immediately
practical. We are also slow. As a result, we are getting left behind.

\textit{Ryan}: Are there any solutions?

\textit{Laird}: Well, certainly we need to make sure that any methods we
publish are accompanied by public-access software that can be used to
replicate the results and apply the methods easily in other settings.
Software also needs to be well documented. Some people do this really well,
but most don't. Of course, it is a double-edged sword since there is not a
lot of quality control. Also, it is difficult and time consuming to write
good software.

\textit{Ryan}: Is there anything we can do to make things easier?

\textit{Laird}: It would be good to close the gap between classical
academic success and contributions more broadly. Rigorous work in
statistics is expected for promotion in most academic departments, but that
work may have little relevance outside of academia. We need to broaden our
criteria for success so that people whose work has been in leadership in
applying statistics at a broader level can become involved in academia, and
vice-versa.

\textit{Ryan}: But isn't that a bit tricky? People can grow a very long CV
comprising second and higher authorship papers, but without having the kind
of leadership I~think you are talking about.

\textit{Laird}: We should be training our students to be leaders and
encouraging them to think about first authorship papers in subject matter
areas. Christl Donnelly, Steve Horvath, Nick Horton, Steve Lake, Fong
Wang-Clow and Bob Glynn are just a few examples of my former students that
come to mind who have had creative, out-of-the-box careers.

\textit{Ryan}: So how do we do that? Should we be training people
differently?

\textit{Laird}: It is hard to train students broadly and deeply at the same
time. They need so much, so it's a question of trade-offs. But something
needs to give because the way we are teaching now, our students get
channeled into a very narrow base. That is OK when career lifetimes are
long and we have opportunity for growth and change after the Ph.D. But in
academia we push people so strongly to do great work in only a few years.
It does not promote diversity of thinking. I~also think computing is really
important. It is a rare person who can do both computing and statistics,
but we should be at least trying to produce such people. We need to train
our students to think algorithmically: how would I~compute that? They also
need to understand data and data structures. Perhaps first and foremost,
they need to know who will be reading their papers and why, and who will be
using their methods.

\textit{Ryan}: I~love that phrase, ``think algorithmically.'' I~think you
are right on target here, Nan. You've set a great example, with so much of
your work either directly focused on algorithms or providing analysis
strategies that lend themselves to effective computation. In fact, you've
had a stellar career, seemingly to imagine the future and push statistics
in the right directions. How did you manage that? Do you have any final
thoughts or advice for young people starting out?

\textit{Laird}: I~think it is really important to like what you do, and
enjoy your work. If your job requires you to do things that do not bring
you pleasure, you are not going to be successful. If you want to write
research papers, you need to enjoy doing the research and writing the
papers. If you want to be a successful teacher, you have to enjoy working
with the students. So many young people ask me what you need to do to
succeed. The really hard question is how to find something that you love
doing. That you have to answer for yourself, but it is what you need to do
if you are to succeed.

\textit{Ryan}: Nan, it's been a pleasure and a privilege to have this
conversation with you. Thanks for your time.\vspace*{12pt}

%\begin{appendix}
%\section{}
%\end{appendix}

% zodis "Acknowledgments" paliekamas pagal autoriu
%\section*{Acknowledgments}

%\begin{supplement}[id=suppA]
%\sname{Supplement A}
%\stitle{}
%\slink[doi]{10.1214/00-STSXXXXSUPP} %[doi,text={...}] - jei reikia
%suskaldyti doi
%\sdatatype{.pdf}
%\sfilename{stsXXXX\_supp.pdf}
%\sdescription{}
%\end{supplement}

% imsref loaded by linak, 2015-09-04 15:42:38
% imsref loaded by linak, 2015-09-04 15:47:20

\end{document}